\documentclass{article}

\usepackage{microtype}
\usepackage{graphicx}
\usepackage{subfigure}
\usepackage{booktabs} 
\usepackage{placeins}
\usepackage{hyperref}

\usepackage[accepted]{icml2024}

\usepackage{amsmath}
\usepackage{amssymb}
\usepackage{mathtools}
\usepackage{amsthm}

\usepackage[capitalize,noabbrev]{cleveref}

\theoremstyle{plain}

\theoremstyle{definition}

\theoremstyle{remark}

\usepackage[textsize=tiny]{todonotes}

\icmltitlerunning{Transfer Learning for Emulating Ocean Climate Variability across $CO_2$ forcing}

\begin{document}

\twocolumn[
\icmltitle{Transfer Learning for Emulating Ocean Climate Variability across CO2 forcing  \\
           }



\icmlsetsymbol{equal}{*}

\begin{icmlauthorlist}
\icmlauthor{Surya Dheeshjith}{equal,yyy}
\icmlauthor{Adam Subel}{equal,yyy}
\icmlauthor{Shubham Gupta}{yyy}
\icmlauthor{Alistair Adcroft}{sch}
\icmlauthor{Carlos Fernandez-Granda}{cds}
\icmlauthor{Julius Busecke}{col}
\icmlauthor{Laure Zanna}{yyy}

\end{icmlauthorlist}

\icmlaffiliation{yyy}{Courant Institute of Mathematical Sciences, New York University}
\icmlaffiliation{cds}{Center for Data Science, New York University}
\icmlaffiliation{col}{Lamont Doherty Earth Observatory, Columbia University}
\icmlaffiliation{sch}{Program in Atmospheric and Oceanic Sciences, Princeton University}

\icmlcorrespondingauthor{Adam Subel}{adam.subel@nyu.edu}

\icmlkeywords{Machine Learning, Ocean, Emulation}

\vskip 0.3in
]



\printAffiliationsAndNotice{\icmlEqualContribution} 

\begin{abstract}
With the success of machine learning (ML) applied to climate reaching further every day, emulators have begun to show promise not only for weather but for multi-year time scales in the atmosphere. Similar work for the ocean remains nascent, with state-of-the-art limited to models running for shorter time scales or only for regions of the globe. In this work, we demonstrate high-skill global emulation for surface ocean fields over 5-8 years of model rollout, accurately representing modes of variability for two different ML architectures (ConvNext and Transformers). In addition, we address the outstanding question of generalization, an essential consideration if the end-use of emulation is to model warming scenarios outside of the model training data. We show that 1) generalization is not an intrinsic feature of a data-driven emulator, 2) fine-tuning the emulator on only small amounts of additional data from a distribution similar to the test set can enable the emulator to perform well in a warmed climate, and 3) the forced emulators are robust to noise in the forcing.

\end{abstract}

\section{Introduction}
\label{introduction}
Recently, emulation for weather and climate models has gone from an emerging field to a resounding success story for how the machine learning community can greatly impact important climate problems. Particularly, we have seen several models surpass ECMWF's state-of-the-art numerical weather models  \cite{price2023gencast,zhong2024fuxi,kochkov2023neural,bi2023accurate,bonev2023spherical}.

The rapid development of emulators has been heavily skewed towards the atmosphere and/or weather timescales, with exciting recent development for atmospheric emulation at longer timescales  \cite{kochkov2023neural,bonev2023spherical,watt2023ace}. There is an emerging interest in the emulation of the ocean, an essential climate component for time scales ranging from years to centuries. Recent works on ocean emulation include time scales of 30 days for global models \cite{xiong2023ai}, seasonal timescales for both idealized and regional ocean modeling \cite{chattopadhyay2023oceannet,bire2023ocean,gray2024long}, and multi-year regional emulation \cite{subel2024building}.

Here we demonstrate the potential of emulation on a global scale for evolving surface ocean fields across multi-year time-scales, while highlighting the accompanying challenges when applying emulators to a changing climate. Using the framework from \citet{subel2024building}, we build emulators forced with atmospheric boundary conditions taken from the climate simulation, which is used as ground truth.

We explore a set of architectures and their ability to skillfully reproduce key metrics from our ground truth model. We then investigate their potential to generalize when providing atmospheric boundary conditions from a warming scenario of the same climate model. While models do not natively extrapolate to distributions far outside the training data, we show that exposure to a small number of samples similar to the test distribution allows the model to generalize well. Finally, we show that these forced emulators are robust to atmospheric  noise. Our results represent a further step forward to help guide the design and evaluation of ocean emulators.

\section{Methods}
\label{method}

The goal is to autoregressively emulate the surface ocean state of a climate model, $ \boldsymbol{\Phi}$, given atmospheric boundary conditions, $ \boldsymbol{F}$, and test the generalization to different atmospheric boundary conditions (for example, taken from climate models with increased CO$_2$ concentrations).

We define the data variables as follows:
\begin{enumerate}
    \item Ocean state $ \boldsymbol{\Phi} = (u, v, T)$: the zonal velocity, meridional velocity, and temperature, respectively, in the surface layer.
    \item Atmosphere boundary conditions $ \boldsymbol{\tau} = (\tau_u, \tau_v, T_{atm})$: the zonal wind stress, meridional wind stress, and air temperature, respectively.
\end{enumerate}

To predict ocean state at a future time step $t+\Delta t$, $\ \boldsymbol{\Phi}_{t+\Delta t}$, we input the ocean state $ \boldsymbol{\Phi}_{t}$ and atmospheric boundary conditions $\boldsymbol{F}_t$ from the current time step $t$. We take $\Delta t = 1~ \mathrm{day}$. This gives 6 input channels and 3 output channels. 

\begin{figure*}[h]
    \centering
    \includegraphics[width=1\linewidth]{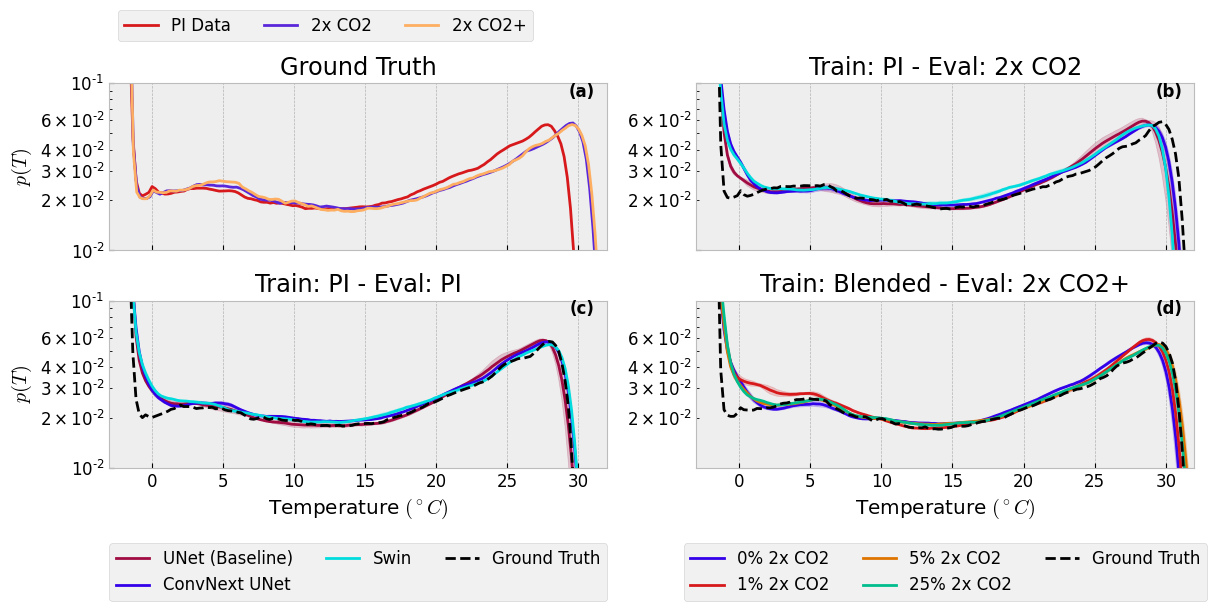}
    \caption{Model skill in reproducing the PDF of temperature.  (a) Comparison of the PDF from model datasets;  (b) Skill for ML models trained on PI and tested on 2xCO2 (out-of-distribution generalization test); (c) Skill for different architectures for models trained and tested on PI data (in-distribution); (d) Transfer learning skill: trained on blended data (PI + some \% of 2xCO2 data) and tested on data from 2xCO2+ run.
    }
    \label{fig:PDFs}
\end{figure*}

\subsection{Data}
\label{data}

We use data from the GFDL CM2.6 coupled climate model, with a horizontal resolution of $1/10^{\circ}$ in the ocean and $1/2^{\circ}$ in the atmosphere \cite{delworth_simulated_2012}.  We conservatively regrid ocean data to a $1^{\circ}$ regular grid, and bilinearly interpolate atmospheric data to the same $1^{\circ}$ regular grid. We use daily data from three CM2.6 runs: 20 years of a preindustrial control (PI) with constant external forcing, 20 years of a transient doubling CO$_2$ experiment sampled from 10 years prior to and 10 years past doubling (2xCO2), and 6 years from a transient quadrupling CO$_2$ experiment taken after the CO$_2$ concentration passes the point of doubling (2xCO2+). The third dataset is only used for testing. The relative sampling windows and CO$_2$ concentrations are shown in figure \ref{fig:co2_scenario}.

We train emulators using 4000 training samples taken daily from the start of the 20 year PI control run. We test on the PI and 2xCO2 runs using an initial state from day 4200  and atmospheric boundary information through day 7200. For the 2xCO2+, we test on the first 2000 days, using day 0 as the initial condition and the remainder for atmospheric boundary information. We train additional emulators using a transfer learning methodology. For such emulators, we take the model trained on PI control data and fine-tune with data from the 2xCO2 run by selecting consecutive samples from the start of the 20-year run (e.g., for 5\% data, we use the first 200 days of the 2xCO2 run).

\begin{figure*}
    \centering
    \includegraphics[width=1\linewidth]{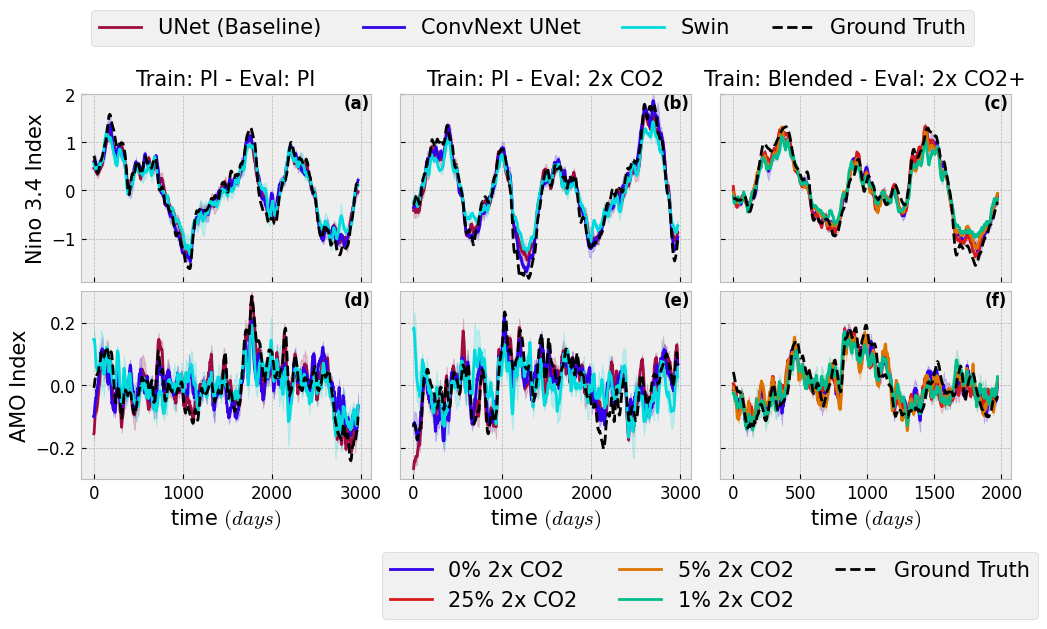}
    \caption{ML Model skill in reproducing key components of climate variability. Panels a-c are for the monthly rolling mean time series of the Nino 3.4 index. Panels d-f for the monthly rolling mean time series of the AMO index. Left and middle columns are ML  models trained on PI control data, and tested on PI or 2xCO2, respectively; right column: tested on blended data  (PI data +  different amount of 2xCO2 data) and tested on 2xCO2+.
}
    \label{fig:Indices}
\end{figure*}

\subsection{Architectures}
\label{archs}
The architectures we use are UNet, ConvNeXT UNet, and Swin Transformer. The models autoregressively predict the ocean states to produce rollouts of any length, provided appropriate boundary conditions are available. All models implement periodic padding along longitude and zero padding at the poles. We briefly describe the ML models below (see Appendix \ref{a:recipe} for further details).

\subsubsection*{UNet}

Our baseline architecture is a UNet, built following \citet{subel2024building}, with encoder and decoder blocks. Each encoder block consists of convolutions and batch normalization layers stacked alternatively. We apply a ReLU activation after each batch normalization layer. The encoder uses max pooling and the decoder uses bilinear upsampling.

\subsubsection*{ConvNeXT UNet}
The ConvNeXT UNet is designed following \citet{subel2024building} and \citet{liu2022convnet}. We replace the encoder blocks with ConvNeXT blocks, which use average pooling and GeLU activation, with blocks that use max pooling and ReLU respectively.

\subsubsection*{Swin Transformer}
We employ the Swin Transformer architecture \cite{liu2021swin}, adapted to produce a large number of pixel-wise outputs, appropriate for our modeling of a dense prediction task. This is built as an encoder-decoder network in a similar fashion to the UNets. Here we start with the ConvNeXT UNet model and replace the encoder with a standard Swin Transformer.

\subsection{Loss Function}

For training the network, we perform multi-step predictions to create a loss function that captures dynamics beyond the time step of the emulator, $\Delta t = 1$ day. For convenience, we use the following notation for recurrent passes of the network: $\tilde{\boldsymbol{\Phi}}_{t+n\Delta t} = \mathcal{F}_\theta^{(n)}(\boldsymbol{\Phi}_t,\boldsymbol{\tau}_t)$, where $^{(n)}$ indicates the number of recurrent passes, $\tilde{\boldsymbol{\Phi}}$ is a predicted state, and $\mathcal{F}_\theta$ is the neural network with parameters $\theta$.
The loss function optimized is given by
\begin{equation}
\mathcal{L}_{\mathrm{mse}}= \sum_{n=1}^{N}{\left \Vert \boldsymbol{\Phi}_{t+n\Delta t} - \mathcal{F}_\theta^{(n)}(\boldsymbol{\Phi}_t,\boldsymbol{\tau}_t) \right \Vert_2^2}
\end{equation}
\noindent Here, $\mathcal{L_{\mathrm{mse}}}$ is the total MSE loss function, $N=4$ is the total number of recurrent passes.

\section{Results}

We use a set of key metrics to capture the skill of the emulators, based on metrics traditionally used for evaluating numerical and statistical models \cite{latif1998review}.
We focus on multi-year time-scales, evaluating the following metrics: probability distributions of state variables (Fig. ~\ref{fig:PDFs}),  representations of key climate indices (Fig. ~\ref{fig:Indices}), and the patterns of bias over multi-year rollouts (Fig. ~\ref{fig:Biases}). Tables with skill scores across architectures, metrics, and different training and testing experiments are given in the appendix.

\subsection{In-Distribution Skill}

The trained ML models skillfully reproduce the probability distribution (PDF) of temperature when trained and tested on PI data (Fig.~\ref{fig:PDFs}c). Our leading model, ConvNext, reproduces the bulk of the PDF well for temperatures warmer than $1^{\circ}\mathrm{C}$. The Swin Transformer has a similar skill to the ConvNext, but the baseline UNet poorly captures the temperature distribution. All models fail to reproduce the near 0$^\circ C$ temperature distributions and create strongly negative, below-freezing temperatures.  This may potentially be alleviated with additional training data (including sea-ice concentration as input, for example) or enforcing an equation of state in future emulators  (e.g., adding salinity as a state variable).

We consider two climate indices of dominant ocean signals to further quantify the model skill on interannual timescales (Fig.~\ref{fig:Indices} panels a and d).
The first index is the Nino 3.4 index, which measures the dominant mode of climate variability and is well captured by all ML models (correlation above .97). This indicates that ML models can respond appropriately to the imposed atmospheric boundary conditions.  The second index is the Atlantic Multidecadal Oscillation (AMO), which is more challenging for the emulators to capture as it may involve deep ocean processes not resolved by our emulator. We still find that our overall best-performing model (ConvNext) correlates above .75 with the ground truth.

The structure of the climatological bias, i.e. the difference in the mean states of the model over a multi-year rollout, shows the error that accumulates over years. All ML models exhibit some biases in the in-distribution tests, and this is particularly evident in Tropics, which has too low kinetic energy for all emulators (Fig.~\ref{fig:bias_PI_PI}b-d). However, these biases are all small ($O( 10) ~ \mathrm{J/m^2}$) relative to the mean state, which is $O( 10^3) ~ \mathrm{J/m^2}$ in the tropics. We show the comparison across architectures in the appendix (Fig.~\ref{fig:bias_PI_PI}).

\subsection{Generalization to a warmer climate}

One of the use cases for ML emulators is to generate realistic long-term trajectories for externally forced runs. To understand the outstanding challenges in generalizing from a stationary system to a different climate, we evaluate our ML models, trained on the PI run, on a warmer climate given an atmosphere from the 2xCO2 run.

All three emulators fail to reproduce the true PDF of the 2xCO2 model (Fig.~\ref{fig:PDFs}b). 
The ConvNext and Swin shift towards the true PDF, with the ConvNext model closing most of the gap.  However, all models fail to capture the warmed range of temperatures, reflected in the bias maps (Appendix, Fig.~\ref{fig:bias_PI_2x}f-h), where there is a uniform global cold bias. A few additional regional biases are present, such as a local cold bias in the Arctic and a warm bias in the North Atlantic and near Antarctica.

Despite the climatological biases, the emulators can reproduce the appropriate variability for the Nino 3.4 and AMO indices. This demonstrates that although the emulators do not capture the mean changes in a warmer climate, they respond to the out-of-distribution atmospheric forcing without becoming unstable or losing track of important atmosphere-forced processes.

In the appendix, we include results that shows the sensitivity of various emulators trained and tested across different datasets and forced with different boundary conditions from the climate model, but also with a uniform $1^\circ \mathrm{C}$ cooling and warming of surface temperature.

\subsection{Transfer Learning: Utilizing Data Across Climates}
To improve the ability of a model trained on the PI run in generalizing to the 2xCO2+ run, which is similar to the 2xCO2 run, we make use of ideas from transfer learning \cite{subel2023explaining,hu2021deep}.  Here, we fine-tune the emulator built on PI data using small amounts of data from the 2xCO2 case.  We explore the requirements on the amount of data, fine-tuning our ConvNext model using 1 (40 samples), 5 (200), and 25\% (1000) of the PI samples used to train the model.

\begin{figure*}
    \centering
    \includegraphics[width=1\linewidth]{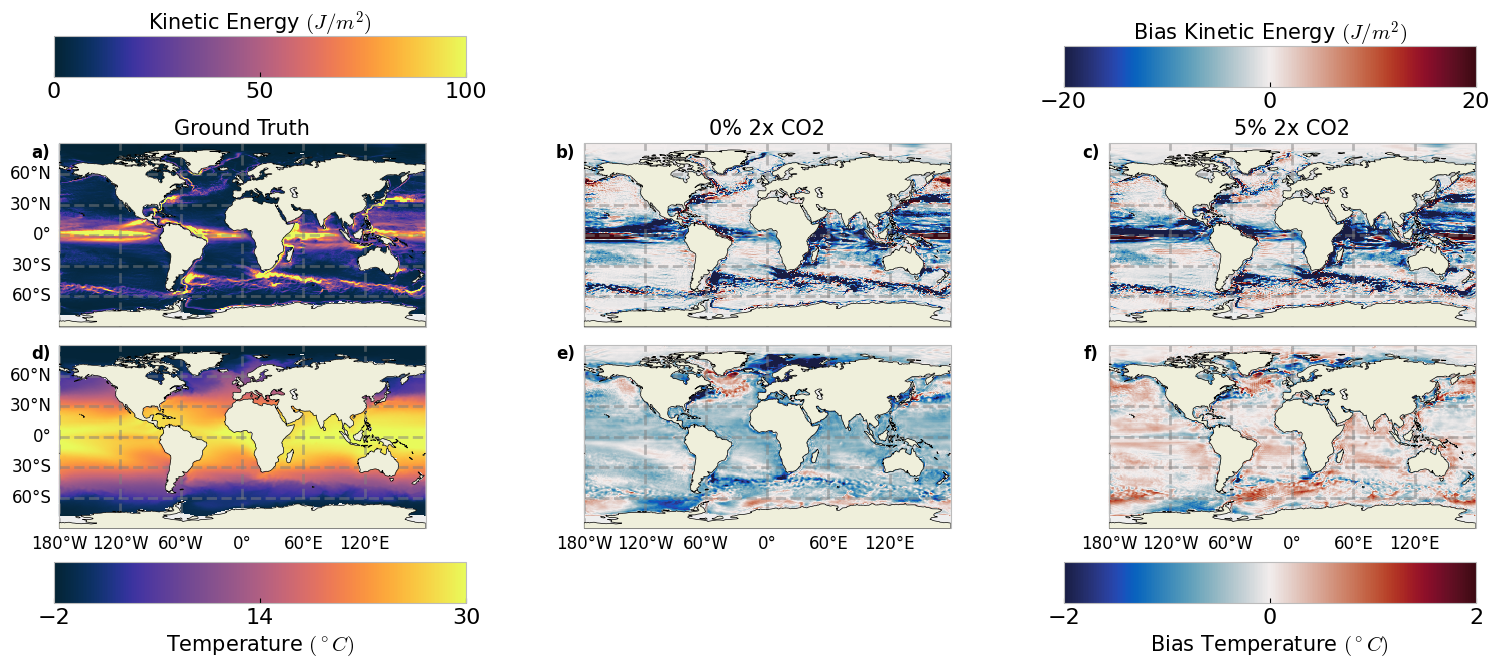}
    \caption{Bias maps (ConvNext prediction $-$ true2xCO2+) for climatological mean for surface kinetic energy (top) and surface ocean temperature (bottom). Panel a and d are the 2xCO2+ ground truth. Training with PI data (b, e), PI + 5\%CO2 (c, f). }
    \label{fig:Biases}
\end{figure*}

The uniform cold bias disappears, even after retraining on only 40 samples from the 2xCO2 run (Appendix Fig.  \ref{fig:Biases}h), though some warm biases emerge in the Southern Ocean. Increasing to 5\% (Fig.  \ref{fig:Biases}e) and then 25\% of additional data yields a major improvement in emulator fidelity, with bias shrinking for each increase in data. 
We obtain a similar behavior for the PDF of temperature, which moves closer to the 2xCO2+ ground truth as the amount of data used for retraining increases (Fig. ~\ref{fig:PDFs} d); using 1\% of additional data, we lose skill at lower temperatures, potentially due to overfitting on a small dataset. As in the other experiments, these models accurately reproduce variability in the form of the Nino 3.4 and AMO index.

Though these results require training data from a distribution similar to the test case, we show that the data burden is quite small when leveraging the training done on the unforced scenario.

\subsection{Robustness to Noisy Boundary Data}

Another use for the emulators is to couple them to multiple components of climate models, and as such, errors will be introduced as the system evolves. We explore the robustness of our emulators to atmospheric noise by introducing  Gaussian noise at each time step during rollout.
The noise is drawn from normal distributions of the form $\mathcal{N}(0,\epsilon \sigma_{\mathbf{F}})$, for values of $\epsilon = $ .05, .25, and 1. We both train and test a ConvNext Unet on the PI run.

\begin{figure*}
    \centering
    \includegraphics[width=0.95\linewidth]{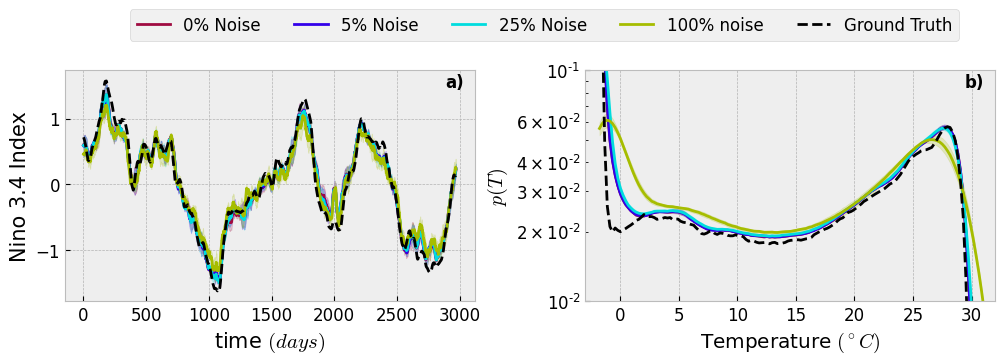}
    \caption{
The impact of atmospheric Gaussian noise (0\%, 5\%, 25\%, 100\%) on the ConvNext emulator's skill. (a) Skill of Nino 3.4 index  (b) The PDF of temperature.
    }
    \label{fig:Noised}
\end{figure*}

We find that the emulator is resilient to noise in the data, with no significant loss of performance at 5\% ($\epsilon=.05$) or even a high value of noise (25\%). In both these tests, the key indices of climate variability remain well represented, and the PDFs remain similar to the noise-free rollout (Fig. ~\ref{fig:Noised}a).

We further increase noise to match the standard deviation of the boundary terms. A large bias is introduced in temperature and kinetic energy, and the PDFs no longer resemble the ground truth (Fig. ~\ref{fig:Noised}a). However, signals much larger than the local standard deviation remain in the emulator rollout. Specifically, the noised emulator reproduces the Nino 3.4 and AMO indices with minimal degradation compared to the cases with less noise added(Fig. ~\ref{fig:Noised}b).

\section{Conclusion and Future Work}
To make machine learning (ML)-based emulators a useful tool for assessing the impacts of climate change, we need an emulator that performs well across metrics on a stationary climate but also under the many possible warming scenarios the future might bring. This work demonstrates the potential of a range of ML models for this problem and examines the potential pitfalls when using a model to generalize far outside the training distribution.

We show that our emulators reproduce key features of climate variability, Nino 3.4 and the AMO index, for both in and out of distribution rollouts. However, when testing the generalization from PI data, the model exhibits large biases and fails to faithfully recreate the temperature PDF. To remedy this problem, we propose a transfer learning approach that utilizes a relatively small sample of data from a warming scenario to significantly improve the generalization of the emulator. We hypothesis that the methodology will apply to any changes in climate regime (e.g., cold and warm paleoclimates).

To couple ocean emulators to either numerical or data-driven models of other climate system components, we need to ensure that small errors in a boundary input do not drive our models to produce unrealistic outputs. We demonstrate that our best-performing emulator retains skill for noisy boundary variables with up to .25 times the standard deviation of those inputs added at each time step. Though there is clearly room to grow in scaling up data and model size, we provide further evidence that the simple framework proposed in \cite{subel2024building} and extended here is a well-founded approach for emulating the ocean from multi-year to decadal time-scales.

\subsubsection*{Acknowledgments}
We thank the M$^2$LInES team for feedback and discussions. We acknowledge NOAA and GFDL for the model data used to perform experiments. This material is based upon work supported by the National Science Foundation Graduate Research Fellowship under Grant No. (DGE-2234660). This project is supported by Schmidt Sciences, LLC.

\newpage
\bibliography{references}
\bibliographystyle{icml2024}
\newpage
\appendix
\section{Training Recipes}
\label{a:recipe}
We train all models on an HPC cluster, using 150GB RAM and 2 NVIDIA RTX 8000s. All models are trained for 3 hours on a batch size of $16$, using an Adam optimizer with a learning rate of $2e-4$, and a Cosine scheduler.

Here, we will further describe the hyper-parameters used to train our models.
\subsection{UNet}
The UNet \cite{subel2024building} has the following channel widths $[64, 128, 256, 512]$ with dilation rates for convolution layers of $[1, 1, 1, 1]$ and
number of layers set to $[2, 2, 2, 2]$. The architecture has a total of $11,813,571$ trainable parameters.
\subsection{ConvNeXT UNet}
The ConvNeXT blocks we use are based on \cite{karlbauer2023advancing} and are modified versions of those described in \cite{liu2022convnet}. \cite{karlbauer2023advancing} do not employ several ConvNeXT features such as large $7 \times 7$ kernels or depthwise separable convolutions. Avoiding these features helps manage the significant increase in parameters and computational load.

The ConvNeXT UNet has channel widths of $[24, 45, 90, 180]$ with dilation rates for convolution layers of $[1, 2, 4, 8]$ and number of layers set to $[1, 1, 1, 1]$. The architecture has a total of $15,887,031$ trainable parameters.

\subsection{Swin Transformer}
The Swin Transformer uses a patch size of $4$ and an embedding dimension of $60$. The number of attention heads for each layer were set to $[3, 6, 10, 15]$ and depth to $[2, 2, 2, 2]$. We use a window size of $10$ and drop path rate of $0.2$.

We address the patching artifacts generated by the embedding layer of a transformer, as seen in \cite{nguyen2023climax} by utilizing a convolutional decoder. Thus, for the decoder, We reuse the core block of ConvNeXT UNet with transposed convolutions instead of bilinear interpolation. The dilation rates were set to [1, 2, 4, 8] and number of layers set to [1, 1, 1, 1]. The architecture has a total of 64, 242, 851 trainable parameters.

\section{Metrics Tables}
We quantify model skill by computing the correlation (Corr) and root mean square error (RMSE) over the time series or mean state of temperature (T), kinetic energy (KE), and climate variablity indices. In Table \ref{tab:model_metrics}, we present statistics for training and evaluating ML models on the PI dataset. Table \ref{tab:new_model_metrics} showcases statistics for training on PI dataset and evaluating on 2xCO2. Table \ref{tab:co2_metrics} presents the statistics for transfer learning with varying amounts of 2xCO2 data evaluated on the 2xCO2+ run. Table \ref{tab:noise_metrics} shows the impact of adding different amounts of noise to the atmosphere boundary conditions.

\begin{table*}
\centering
\begin{tabular}{|l|c|c|c|c|c|c|c|c|}
\hline
Model Name      & KE Corr  & KE RMSE   & T RMSE   & PDF T Corr & Nino Corr & Nino RMSE & Amo Corr  & Amo RMSE \\ \hline
UNet (Baseline) & 0.921& 20.462& 0.391& 0.812& 0.988& 0.132& 0.917& 0.026
\\ \hline
ConvNext UNet   & 0.932& 18.024& 0.334& 0.833& 0.983& 0.116& 0.775& 0.043
\\ \hline
Swin            & 0.942& 16.559& 0.441& 0.846& 0.972& 0.176& 0.742& 0.044
\\ \hline
\end{tabular}
\caption{Emulator Statistics when training and evaluating on the PI dataset. Note the temperature correlation and the kinetic energy PDF correlation  are removed as all architectures have a value above .99.}
\label{tab:model_metrics}
\end{table*}

\begin{table*}
\centering
\begin{tabular}{|l|c|c|c|c|c|c|c|c|}
\hline
Model Name      & KE Corr  & KE RMSE   & T RMSE   & PDF T Corr & Nino Corr & Nino RMSE & Amo Corr  & Amo RMSE \\ \hline
UNet (Baseline) & 0.907& 21.334& 0.617& 0.941& 0.98& 0.174& 0.85& 0.035
\\ \hline
ConvNext UNet   & 0.913& 19.526& 0.641& 0.888& 0.982& 0.133& 0.822& 0.038
\\ \hline
Swin            & 0.919& 20.663& 0.928& 0.9& 0.971& 0.252& 0.402& 0.064
\\ \hline
\end{tabular}
\caption{Emulator Statistics when training and evaluating on the PI dataset and evaluating on 2xCO2. Note the temperature correlation and the kinetic energy PDF correlation  are removed as all architectures have a value above .99.}
\label{tab:new_model_metrics}
\end{table*}

\begin{table*}
\centering
\begin{tabular}{|l|c|c|c|c|c|c|c|c|}
\hline
Model Name   & KE Corr  & KE RMSE   & T RMSE   & PDF T Corr & Nino Corr & Nino RMSE & Amo Corr  & Amo RMSE \\ \hline
0\% 2xCO2  & 0.907& 19.6& 0.628& 0.895& 0.978& 0.146& 0.917& 0.027
\\ \hline
1\% 2xCO2  & 0.846& 25.604& 0.52& 0.766& 0.972& 0.216& 0.861& 0.035
\\ \hline
5\% 2xCO2  & 0.904& 20.658& 0.401& 0.855& 0.976& 0.18& 0.864& 0.033
\\ \hline
25\% 2xCO2 & 0.931& 18.123& 0.367& 0.809& 0.98& 0.136& 0.86& 0.035
\\ \hline
\end{tabular}
\caption{Emulator Statistics when varying the amount of data taken from the 2xCO2 to retrain the ConvNext model trained on PI. This is evaluated on the 2xCO2+ data. Note the temperature correlation and the kinetic energy PDF correlation  are removed as all architectures have a value above .99.}
\label{tab:co2_metrics}
\end{table*}

\begin{table*}
\centering
\begin{tabular}{|l|c|c|c|c|c|c|c|c|}
\hline
Model Name      & KE Corr  & KE RMSE   & T RMSE   & PDF T Corr & Nino Corr & Nino RMSE & Amo Corr  & Amo RMSE \\ \hline
100\% Noise     & 0.907& 24.371& 0.652& 0.378& 0.985& 0.15& 0.775& 0.043
\\ \hline
25\% Noise      & 0.933& 18.125& 0.337& 0.704& 0.984& 0.114& 0.789& 0.042
\\ \hline
5\% Noise       & 0.933& 17.956& 0.333& 0.824& 0.984& 0.116& 0.794& 0.042
\\ \hline
 0\% Noise       & 0.932& 18.024& 0.334& 0.833& 0.983& 0.116& 0.775&0.043
\\ \hline
\end{tabular}
\caption{Emulator Statistics when training and evaluating on the PI dataset with noise added to the atmosphere boundary conditions. Note the temperature correlation and the kinetic energy PDF correlation  are removed as all architectures have a value above .99.}
\label{tab:noise_metrics}
\end{table*}

\section{Forcing Comparisons}
\begin{figure}
    \centering
    \includegraphics[width=1\linewidth]{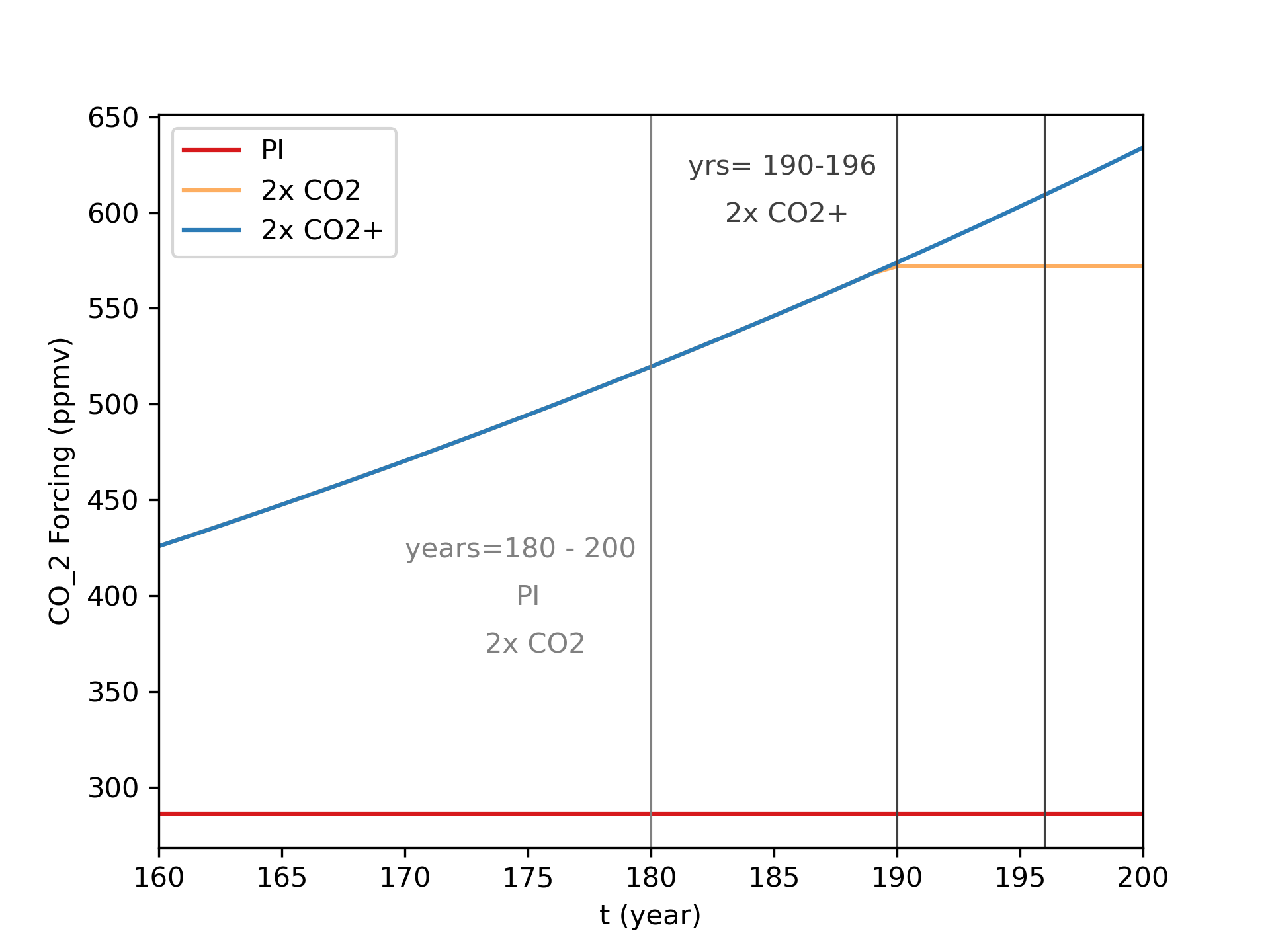}
    \caption{Comparison of the CO$_2$ trajectories within the model runs used for this work. The PI run and 2x CO2 run take data from model years 180 to 200. For the 2x CO2 run, this corresponds to 10 years of incremental increase to the doubling point and 10  years of stationary forcing past doubling. For the 2x CO2+ run, the 6 years are years 190 through 196, which are years with incremental increase past the doubling point.}
    \label{fig:co2_scenario}
\end{figure}

 To give better context to the difference between the scenarios in each run, figure \ref{fig:co2_scenario} shows the CO2 forcing as a function of model year across scenarios. In addition in indicates the range of years included from each run.

\section{Additional Bias Figures}

\begin{figure*}
    \centering
    \includegraphics[width=1\linewidth]{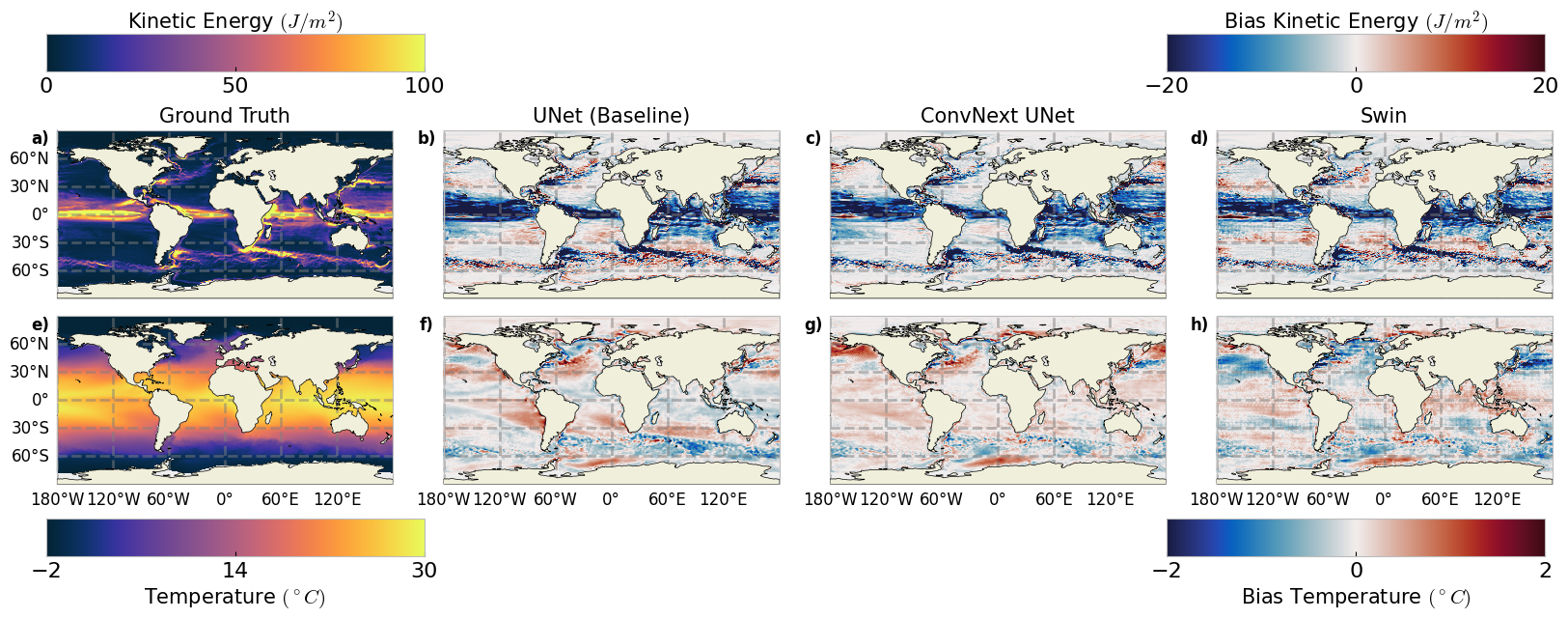}
    \caption{Bias maps (train PI run and test PI run) for climatological mean for surface kinetic energy (top) and surface ocean temperature (bottom). Panel a and e are the PI ground truth. Baseline UNet (b, f), ConvNext  (c, g), Swin (d, h). }
    \label{fig:bias_PI_PI}
\end{figure*}

\begin{figure*}
    \centering
    \includegraphics[width=1\linewidth]{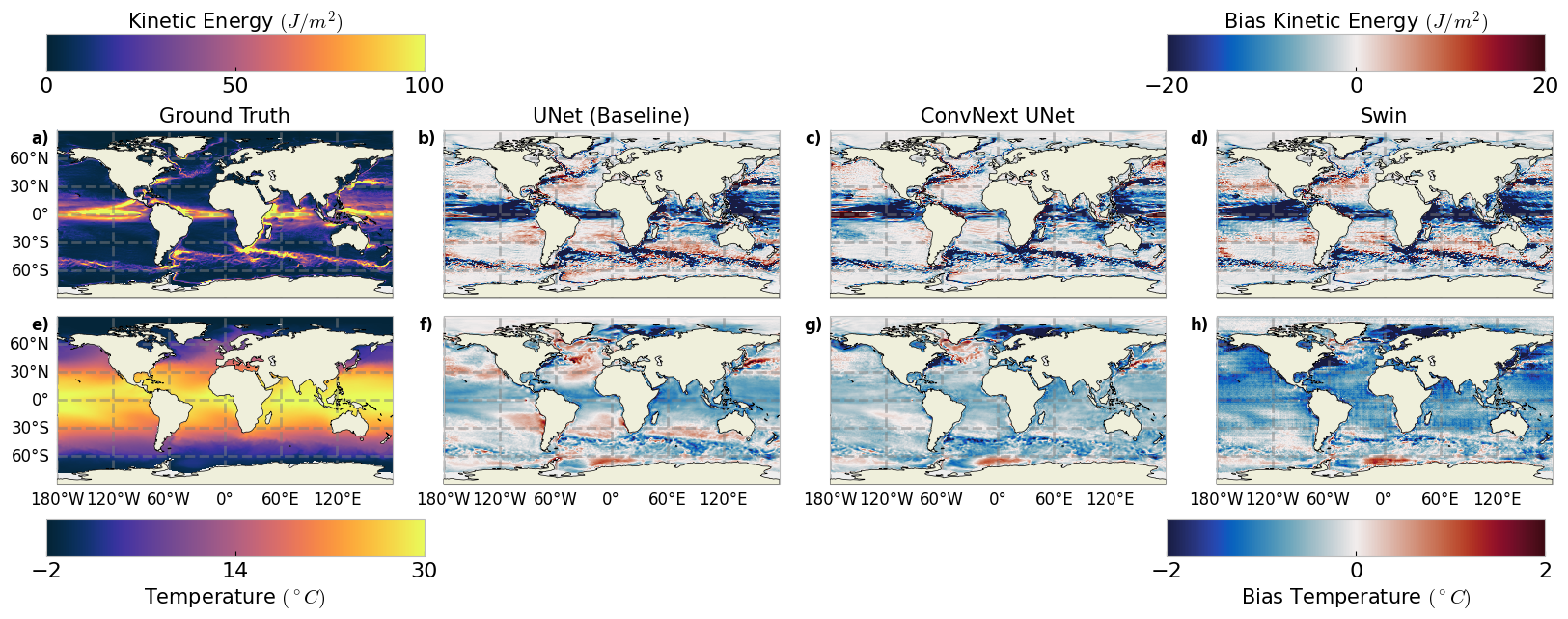}
    \caption{Bias maps (train PI run and test 2x CO2 run) for climatological mean for surface kinetic energy (top) and surface ocean temperature (bottom). Panel a and e are the PI ground truth. Baseline UNet (b, f), ConvNext  (c, g), Swin (d, h). }
    \label{fig:bias_PI_2x}
\end{figure*}

Here we show additional bias plots to complement the results in the main text. Figure \ref{fig:bias_PI_PI} shows the bias training and testing on the PI run and figure \ref{fig:bias_PI_2x} shows the bias for each architecture when training on the PI run and generalizing to the 2xCO2 run. We also include the extended version of figure \ref{fig:Biases} that includes all retraining percentages.

\begin{figure*}
    \centering
    \includegraphics[width=1\linewidth]{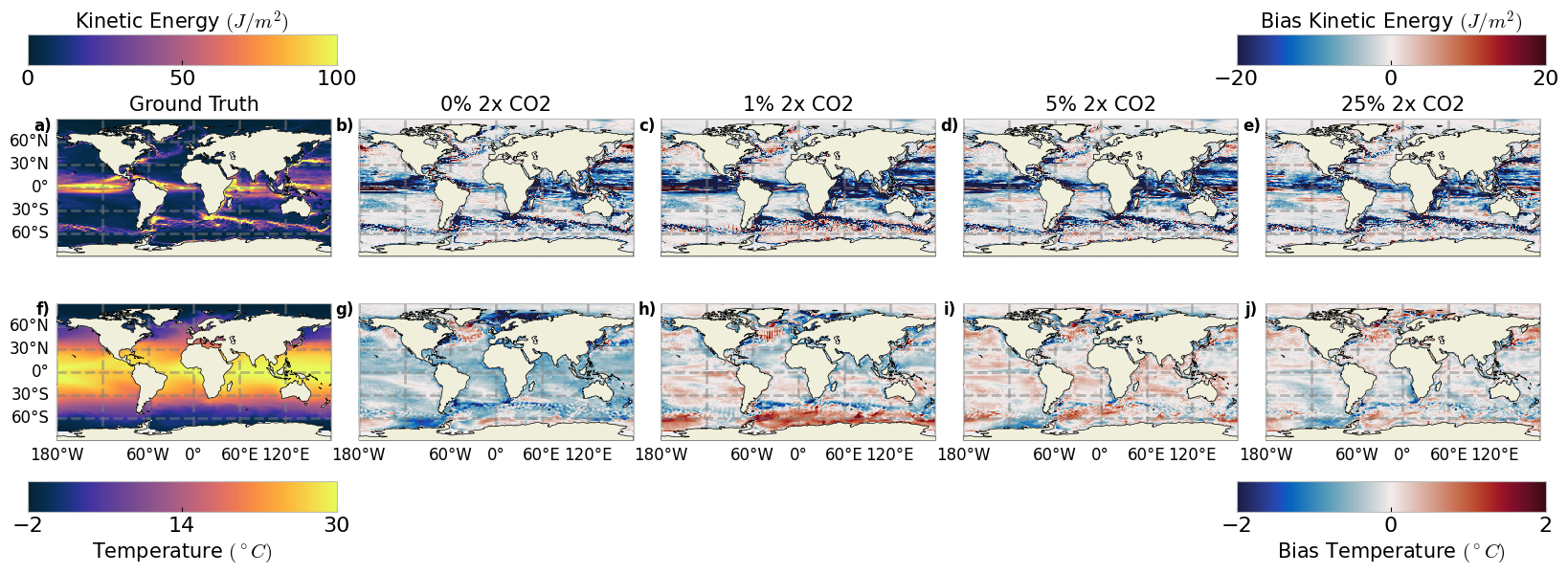}
    \caption{Bias maps (ConvNext prediction $-$ true2xCO2+) for climatological mean for surface kinetic energy (top) and surface ocean temperature (bottom). Panel a and f are the 2xCO2+ ground truth. Training with PI data (b, d), PI + 1\%CO2  (c, h), PI + 5\%CO2 (d, i), PI + 25\%CO2 (e, j).
    }
    \label{fig:Biases_full}
\end{figure*}

\section{Time Series Plots}
\begin{figure*}
    \centering
    \includegraphics[width=1\linewidth]{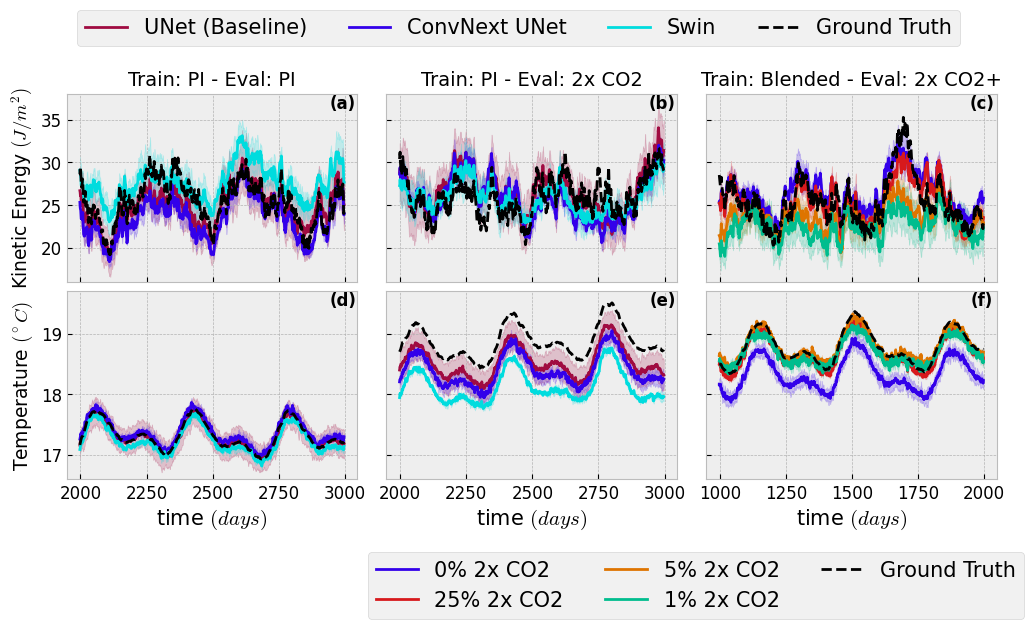}
    \caption{ML Model skill in reproducing time series of model state variables. Panels a-c are for the time series of the global mean kinetic energy. Panels d-e for the time series of the global mean temperature. The left and middle columns are ML  models trained on PI control data and tested on PI or 2xCO2, respectively; the right column is tested on blended data  (PI data +  different amounts of 2xCO2 data) and tested on 2xCO2+. }
    \label{fig:timeseriesappend}
\end{figure*}

In Figure \ref{fig:timeseriesappend}, we present the ability of the models to reproduce the global mean time series of kinetic energy and temperature across the different training settings.

\section{Additional noised results}
\begin{figure*}
    \centering
    \includegraphics[width=1\linewidth]{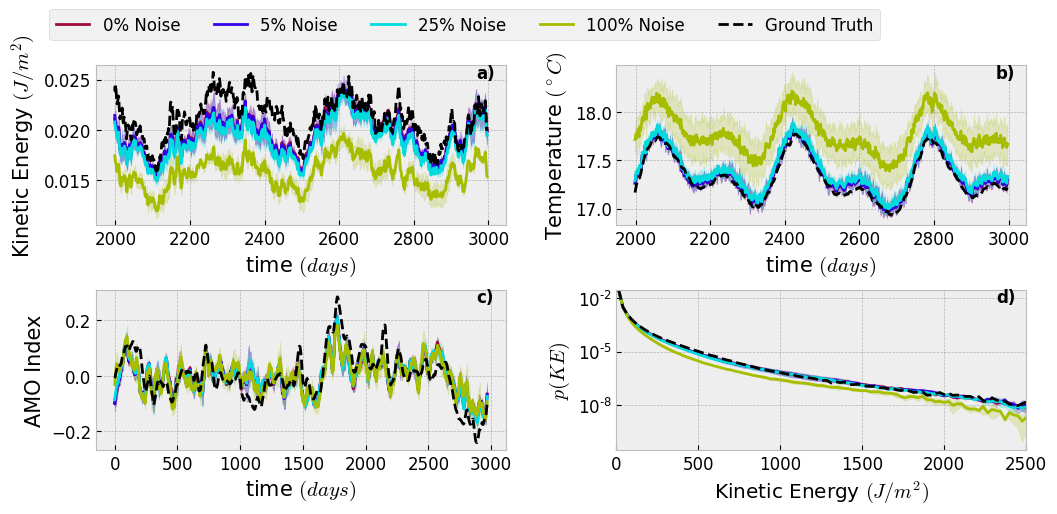}
    \caption{The impact of atmospheric Gaussian noise (0\%, 5\%, 25\%, 100\%) on the ConvNext emulator's skill. (a) kinetic energy time series (b) temperature time series (c) Skill of AMO index  (d) The PDF kinetic energy. }
    \label{fig:noise_append}
\end{figure*}

\begin{figure*}
    \centering
    \includegraphics[width=1\linewidth]{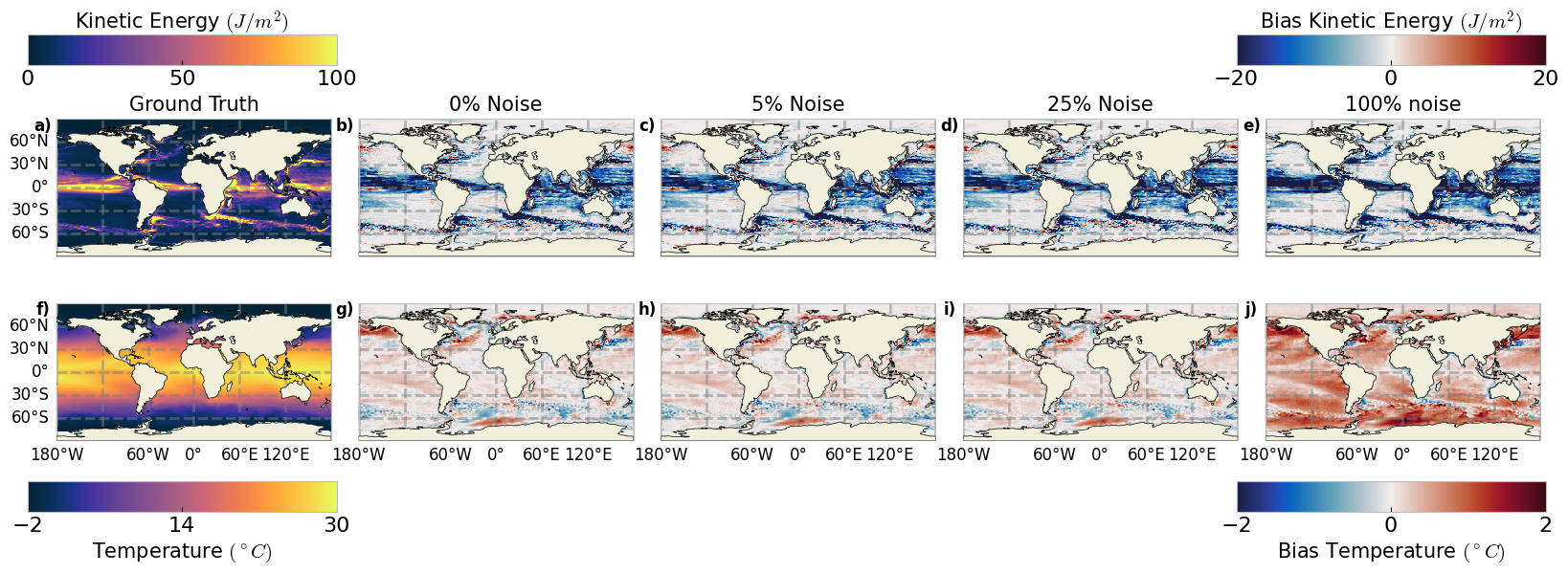}
    \caption{Bias maps (train PI run and test PI with varying levels of Gaussian noise) for climatological mean for surface kinetic energy (top) and surface ocean temperature (bottom). Panel a and f are the PI ground truth. Noise free (b, g), 5\% noise  (c, h),  25\% noise  (d, i), 100\% noise (e,j). }
    \label{fig:bias_noise}
\end{figure*}

To complement the results of figure ~ \ref{fig:Noised}, we include additional metrics in figure ~\ref{fig:noise_append} and bias in figure ~\ref{fig:bias_noise}.

\section{Perturbation Experiments}
\begin{figure*}
    \centering
    \includegraphics[width=1\linewidth]{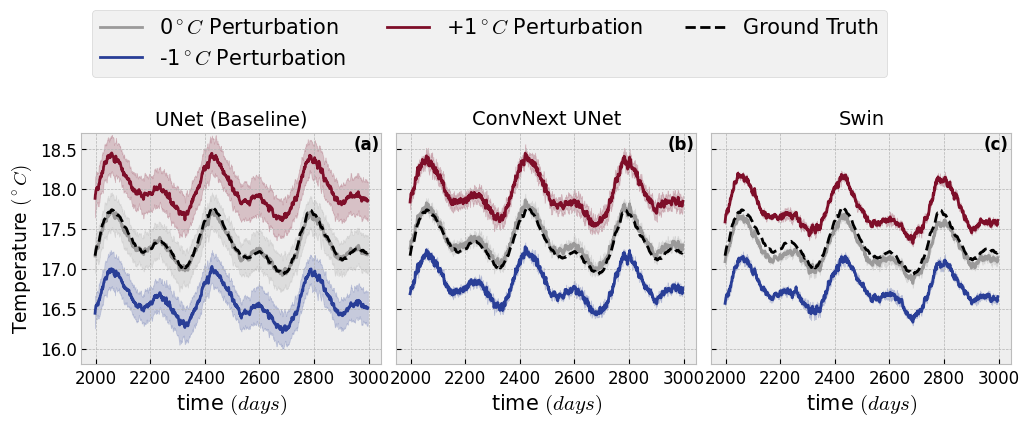}
    \caption{Model sensitivity to perturbations of atmospheric surface air temperature. Each panel shows the sensitivity of a particular architecture through the global mean temperature time series. }
    \label{fig:Perturbation}
\end{figure*}

To ensure that our models are sensitive to a simple uniform perturbation of surface air temperature, we take our models trained on the PI run and evaluate the model with an atmosphere taken from the evaluation window of the PI run, but with 1$^\circ \mathrm{C}$ uniformly added and removed at each time step. In figure \ref{fig:Perturbation}, the models respond well, with a uniform increase and decrease around the mean state.

\end{document}